\documentclass[aps,preprint,superscriptaddress,nofootinbib]{revtex4-1}

\usepackage{amsmath}
\usepackage{amssymb}
\usepackage{amsfonts}
\usepackage{amscd}
\usepackage{accents}
\usepackage{epsfig}
\usepackage{graphicx,amssymb,amsmath,wasysym}

\usepackage{graphicx}

\newcommand{\bmat}{\left(\begin{array}}
\newcommand{\emat}{\end{array}\right)}
\newcommand{\be}{\begin{equation}}
\newcommand{\ee}{\end{equation}}
\newcommand{\ba}{\begin{eqnarray}}
\newcommand{\ea}{\end{eqnarray}}

\def\lsim{\raise0.3ex\hbox{$\;<$\kern-0.75em\raise-1.1ex\hbox{$\sim\;$}}}
\def\gsim{\raise0.3ex\hbox{$\;>$\kern-0.75em\raise-1.1ex\hbox{$\sim\;$}}}

\def\be{\beta}

\def\nn{\nonumber}

\def\beq{\begin{equation}}
\def\eeq{\end{equation}}
\def\bal#1\eal{\begin{align}#1\end{align}}

\def\round#1{\left({#1}\right)}

\newcommand{\lif}{\lambda_\phi}
\newcommand{\muf}{\mu_\phi}
\newcommand{\lih}{\lambda_h}
\newcommand{\lhf}{\lambda_{h\phi}}

\usepackage{color}
\newcommand{\mk}[1]{#1}

\begin{document}

\title{Higgs--Inflaton Mixing and Vacuum Stability}

\author{Yohei Ema}
\affiliation{Department of Physics, Faculty of Science, The University of Tokyo, Bunkyo-ku, Tokyo 113-0033, Japan}

\author{Mindaugas Kar\v{c}iauskas}
\affiliation{University of Jyvaskyla, Department of Physics, P.O.Box 35 (YFL), FI-40014\\ University of Jyv\"{a}skyl\"{a}, Finland}

\author{Oleg Lebedev}
\affiliation{University of Helsinki and Helsinki Institute of Physics, P.O. Box 64, FI-00014, Helsinki, Finland}

\author{Stanislav Rusak}
\affiliation{Nordita, KTH Royal Institute of Technology and Stockholm University, Roslagstullsbacken 23, SE-106 91 Stockholm, Sweden}

\author{Marco Zatta}
\affiliation{University of Helsinki and Helsinki Institute of Physics, P.O. Box 64, FI-00014, Helsinki, Finland}

\begin{abstract}
The quartic and trilinear Higgs field couplings to an additional real scalar are renormalizable, gauge and Lorentz invariant. Thus, on general grounds, one expects such couplings between the Higgs and an inflaton in quantum field theory. In particular, the (often omitted) trilinear coupling is motivated by the need for reheating the Universe after inflation, whereby the inflaton decays into the Standard Model (SM) particles.      
Such a coupling necessarily leads to the Higgs--inflaton mixing, which could stabilize the electroweak vacuum by increasing the Higgs self--coupling. We find that the inflationary  constraints on the trilinear coupling are weak such that the Higgs--inflaton mixing up to order one is allowed, making it accessible to colliders.  
This entails  an exciting possibility  of  a direct inflaton search at the LHC.
\end{abstract}

\maketitle 

\section{Introduction}

The current data favor metastability of the electroweak  (EW) vacuum, although the result is very sensitive to the  top quark mass~\cite{Buttazzo:2013uya,Bezrukov:2012sa,Alekhin:2012py,Bednyakov:2015sca}. Assuming that our vacuum is indeed metastable, we face a number of cosmological  challenges including why the Universe has chosen an energetically disfavored state and why it stayed there during inflation despite quantum fluctuations~\cite{Espinosa:2007qp,Lebedev:2012sy}. Minimal solutions to these puzzles require modification of the Higgs potential during inflation only~\cite{Lebedev:2012sy}, although introduction of a single extra scalar is sufficient to make the electroweak vacuum completely stable~\cite{Lebedev:2012zw,EliasMiro:2012ay}.

In this Letter, we suggest another minimal option
which does not employ any extra fields beyond the usual inflaton. We show that
 the Higgs mixing with an inflaton can lead to a stable EW vacuum. A trilinear Higgs--inflaton coupling always leads to such a mixing and  it is generally present in realistic models describing the reheating stage correctly~\cite{Gross:2015bea}.  We find that cosmological constraints on this coupling are weak
and an order one mixing is possible.
In this case, the model is effectively described by a single mass scale of the EW size
 making it particularly interesting for direct LHC searches.

\subsection{THE SET-UP}

In  quantum field theory, one should include all the couplings that are (up to) 
 dimension--4, gauge and Lorentz invariant. Thus, on general grounds, we expect 
a quartic $H^\dagger H \phi^2$ 
  and a trilinear  $H^\dagger H \phi$    interaction   
between the Higgs field and an inflaton   $\phi$. The   presence of the trilinear term can be motivated 
by the need for reheating the Universe after inflation: the inflaton transfers (at least in part) its energy to the SM particles through decay and  the relevant interactions generate 
the $H^\dagger H \phi$  term at loop level~\cite{Gross:2015bea}. It can only be forbidden if the inflaton is assummed to be stable, for instance, due to the $\phi \rightarrow -\phi$ symmetry, and constitutes part of dark matter~\cite{Lerner:2009xg}. However, it is not clear whether this symmetry remains exact in quantum gravity. 

Apart from the renormalizable QFT interactions, the Higgs dynamics are affected by its
coupling to gravity. Although gravity is non--renormalizable, one may focus on the coupling of lowest dimension  $H^\dagger H  \hat R$~\cite{Chernikov:1968zm}, with $\hat R$ being the scalar curvature, assuming that the effective field theory expansion applies.
 In any case, such a coupling is generated radiatively~\cite{Buchbinder:1992rb}.

 Thus, on general grounds, we expect the following leading interactions between the Higgs 
 and an inflaton/gravity (see also~\cite{Ema:2017loe}),
   \begin{eqnarray}
&&  -{\cal L}_{h\phi } = \lambda_{h \phi} H^\dagger H \phi^2 + 2 \sigma H^\dagger H \phi \,, \nonumber  \\
&& -{\cal L}_{hR } = \xi_h H^\dagger H  \hat R \;.
 \end{eqnarray}
It is interesting that this setup necessarily leads to the mixing between the Higgs and the inflaton.  This is required by the $H^\dagger H \phi$ term with $H$ developing a vacuum expectation value.

Including an analogous $\phi$ coupling to gravity and all renormalizable $\phi$--self-interactions, we obtain the following Jordan frame action:
\begin{equation}
S= \int d^4 x \sqrt{- \hat g} \left[ {1\over 2} \Omega^2 \hat R - {1\over 2}  \hat g^{\mu\nu}
\partial_\mu \phi \; \partial_\nu \phi -  {1\over 2}  \hat g^{\mu\nu} \partial_\mu h \; \partial_\nu h -V(\phi,h)
\right],
\end{equation} 
 where we have set $M_{\rm Pl}=1$ and used the unitary gauge $H= (0,h/\sqrt{2})^T$. 
 The frame function $\Omega^2$ and the potential $V(\phi,h)$ are given by
 \begin{eqnarray}
 && \Omega^2 = 1 + \xi_\phi \phi^2 + \xi_h h^2      \;, \nonumber \\ 
 && V(\phi,h)=  \frac{\lambda_h}{4} h^4 - \frac{\mu_h^2}{2}h^2 +\frac{\lambda_{h\phi}}{2}
h^2 \phi^2 + \sigma h^2 \phi + \frac{\lambda_\phi}{4}\phi^4 + \frac{b_3}{3}\phi^3
- \frac{\mu_\phi^2}{2}\phi^2 + b_1 \phi \,,
\label{pot}
 \end{eqnarray}
 where we have eliminated the $\phi \hat R$ term by field redefinition of $\phi$.
 We take $\lambda_\phi >0$, $\xi_\phi \gg \vert \xi_h \vert, 1$ as well as $\lambda_h >0$ at the inflation scale, which we justify later by the Higgs--inflaton mixing.  Further, we assume that all the dimensionful parameters are far below the Planck scale. In a particularly interesting case of a single mass scale, these parameters are of electroweak size.  
 

\section{Inflation}

 
 In what follows, we consider a representative inflation model which fits the PLANCK data~\cite{Ade:2015lrj} very well. That is, we assume that inflation is driven by the non--minimal $\phi$ coupling to gravity $\xi_\phi \phi^2 \hat R$ with $\xi_\phi \phi^2 \gg 1$, in analogy with the  ``Higgs inflation'' model~\cite{Bezrukov:2007ep}.  
The transition to the Einstein frame, where the curvature--dependent term becomes the usual $ R/2$, is achieved by the metric rescaling~\cite{Salopek:1988qh}
\begin{equation}
 g^{\mu\nu} = \Omega^2 \hat g^{\mu\nu} \;.
\end{equation} 
This induces non--canonical kinetic terms for the scalars. 
Since
  $\vert \xi_h \vert \ll \xi_\phi$, $h \ll \phi$ and   the dimensionful quantities are far below the Planck scale,  during inflation one can neglect all the terms apart from 
 $ \lambda_\phi \phi^4$ and $\xi_\phi \phi^2$.  For the canonically normalized variable $\chi$ satisfying 
 \begin{equation}
{d\chi \over d \phi} = {\sqrt{  1+\xi_\phi (1+6 \xi_\phi) \phi^2} \over 1+ \xi_\phi \phi^2  } \;,
\end{equation} 
 one finds $\chi \simeq \sqrt{3\over 2} \ln \xi_\phi \phi^2 $ in our regime
 and the potential is given by 
\mk{
\begin{equation}
U(\chi) \simeq {\lambda_\phi \over 4 \xi_\phi^2} \left( 1-e^{-\sqrt{2\over 3} \chi} \right)^2\;, 
\end{equation} 
}
\mk{where $U\equiv V/\Omega^4$.} At $\chi \gg 1$,  it is exponentially close to a flat potential and thus supports inflation. The CMB normalization~\cite{Lyth:1998xn} requires $\lambda_\phi / \xi_\phi^2  \simeq 0.5 \times 10^{-9}$. A further constraint on these parameters comes from unitarity considerations. The unitarity cutoff  scale of our theory is given by $\xi_\phi^{-1}$ at which higher dimension operators cannot be ignored~\cite{Burgess:2009ea,Barbon:2009ya}, while the energy density during inflation  is of order 
$ \lambda_\phi/\xi_\phi^2 $. Requiring  $\xi_\phi^{-4} \gsim  \lambda_\phi/\xi_\phi^2  $,
one finds 
 $ \lambda_\phi \xi_\phi^2  \lsim 1.$
 Combining this with the CMB normalization constraint, we get 
\begin{equation}
  \lambda_\phi (\Lambda_I)  \lsim 2 \times 10^{-5}  
  \label{unit}
\end{equation}
and $\xi_\phi (\Lambda_I)  \lsim 2 \times 10^{2}  $, where $\Lambda_I$ is the inflation scale which can be taken to be \mk{$U^{1/4}\sim  (\lambda_\phi/\xi_\phi^2)^{1/4}  $}.\mk{\footnote{For  the renormalization group running of the couplings,  we take
$\Lambda_I \sim M_{Pl}$ to simplify numerical computations.
}}
This may be a somewhat conservative bound~\cite{Bezrukov:2010jz}. 
We further impose the condition that the radiative corrections to the inflaton potential, e.g. in the Coleman--Weinberg form, be small (see, for example, \cite{Lebedev:2012sy}). This gives approximately $\lambda_{ h \phi}^2 / 16 \pi^2  \ll \lambda_\phi$ 
restricting $\lambda_{ h \phi}$ to be below $10^{-2}$ at the inflation scale.
On the other hand, the Coleman--Weinberg correction induced by the trilinear $\phi h^2$
term is negligible: it is suppressed by $(\sigma/\phi)^2$ which is vanishingly small in the range of interest.

During inflation, the Higgs field is a spectator. \mk{For $\lambda_h > 0$ and $\lambda_{h\phi}$ in the range of interest, it is a heavy field at the inflation scale, with mass of order $\sqrt{\lambda_{h\phi} / \xi_\phi} \gg H_I$, 
stabilized at the origin~\cite{Lebedev:2011aq}}.
Since the inflationary dynamics are dictated by the quartic couplings, the Higgs--inflaton
mixing is completely negligible at this stage.

The inflationary predictions of the model are in excellent agreement with the PLANCK data.
In particular, the scalar spectral index is predicted to be $n_s\simeq 0.97$ and the tensor-to-scalar ratio is $r \simeq 3\times 10^{-3}$~\cite{Bezrukov:2007ep}. The latter is within the range of detectability by future CMB missions~\cite{Errard:2015cxa}. Note that, unlike the Higgs inflation scenario, our model is free of significant radiative corrections.

\section{Preheating and reheating}

During inflation the $\chi$ field slowly rolls towards smaller values, while the Higgs
is anchored at the origin by the inflaton--induced effective mass.  
When $\chi$ reaches the critical value $\chi=\chi_{\mathrm{end}}\simeq\sqrt{3/2}\ln\left(1+2/\sqrt{3}\right)$
\cite{GarciaBellido:2008ab}, the slow-roll ends and  $\chi$ rolls fast
  to the minimum of the potential where it oscillates with a decaying
amplitude.

\begin{figure}
\includegraphics[scale=0.3]{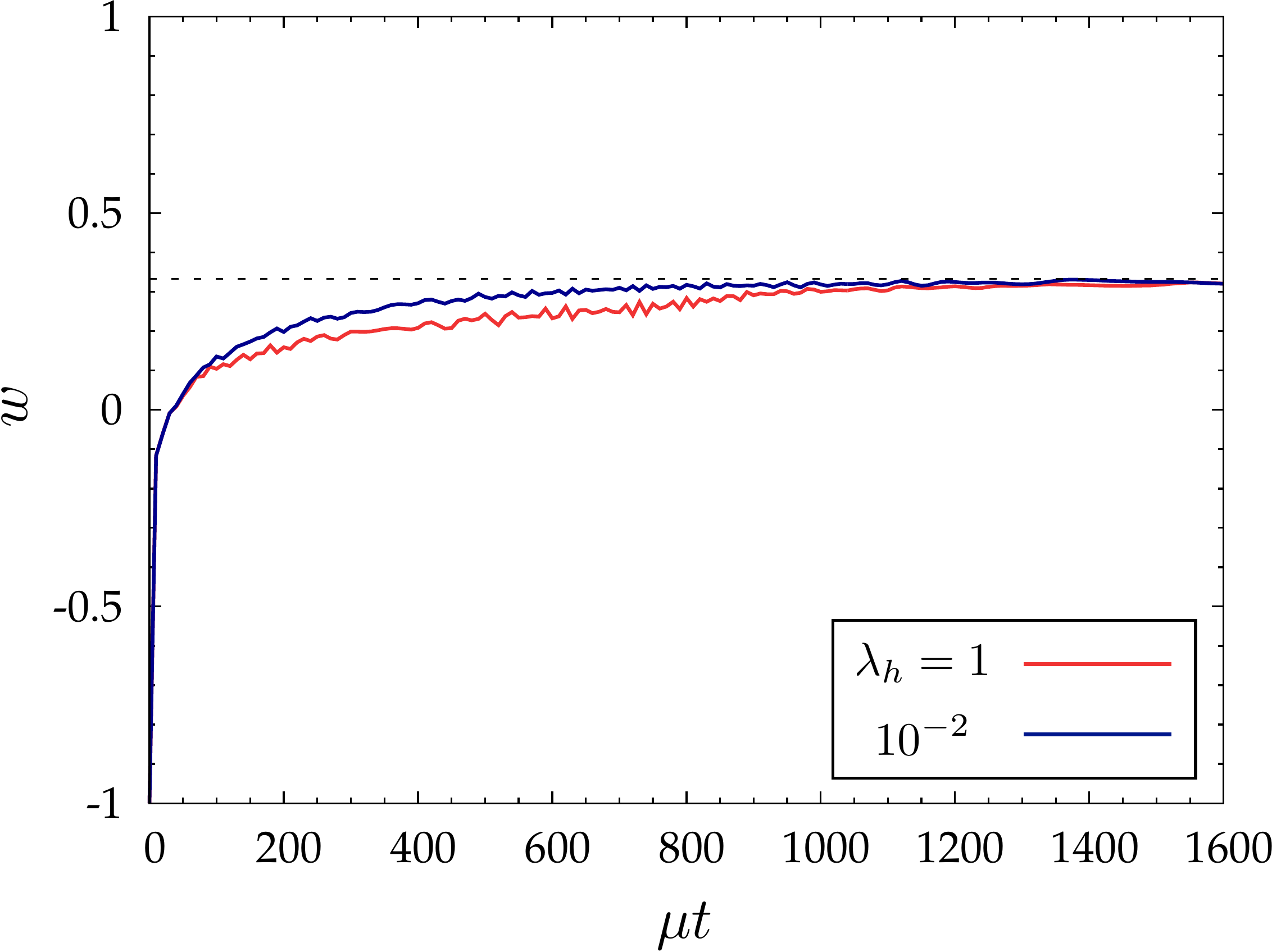} 
\hspace{0.05cm}
\includegraphics[scale=0.3]{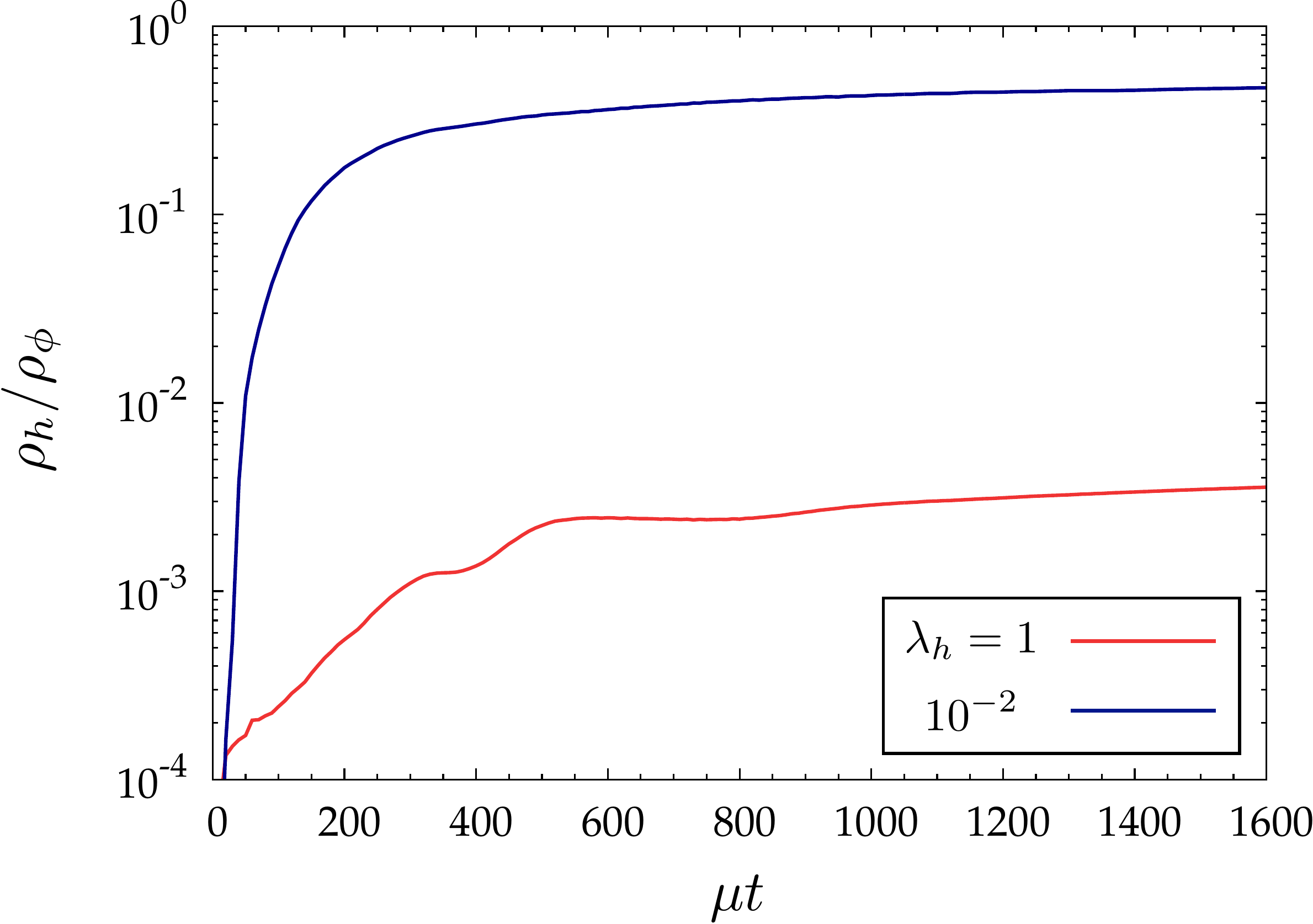} 
\caption{\label{fig:eos}{\it Left:} Evolution of the equation of state for representative values of $\lambda_h (\Lambda_I)$ (lattice simulation). Here  $\mu= \sqrt{\lambda_\phi /3 \xi^2_\phi}$, $\xi_\phi (\Lambda_I) =10^2  $, $\lambda_{h\phi} (\Lambda_I)=10^{-3}$ and at late times $w=p/\rho$ approaches 1/3.
{\it Right:} Ratio between the Higgs and the inflaton energy densities. For $\lambda_h (\Lambda_I)=10^{-2}$, the Higgs quanta are produced efficiently through parametric resonance.}
\end{figure}

In terms of the original variable $\phi$, inflation ends at  $\phi \sim 1/\sqrt{\xi_\phi}$.
As its amplitude descreases further, the relevant for preheating regimes are described by the
canonically normalized inflaton $\chi$ via the relation
\begin{equation}
\chi \simeq \left \{
  \begin{tabular}{ccc}
 $\phi$  &  for & $\phi^2 \ll {1\over 6 \xi_\phi^2 } ~,$ \\
  $\pm \sqrt{3\over 2} \xi_\phi \phi^2$ & for   & ${1\over 6 \xi_\phi^2 } \ll \phi^2 \ll  {1\over  \xi_\phi } ~.$
  \end{tabular}
\right. 
\end{equation}
In these regimes, the potential is $U(\chi) = {1\over 4} \lambda_\phi \chi^4$ and
$U(\chi) = { \lambda_\phi \over 6 \xi_\phi^2}  \chi^2$, respectively.
The inflaton starts oscillating  in the quadratic potential with the effective mass--squared 
$\mu^2 = { \lambda_\phi \over 3 \xi_\phi^2}$, while its amplitude decreases  as $(\mu t)^{-1}$. 
Thus, after $\mu t \sim {\cal O}(6\xi_\phi)$ the system enters the quartic regime and \mk{the inflaton becomes massless (at the classical level)}. 
At this stage, the Universe quickly becomes radiation--dominated~\cite{Lozanov:2017hjm}  although that does not imply thermal equilibrium. In particular,
as shown in the left panel of Fig.~\ref{fig:eos},
 the equation of state approaches that of radiation, $p=w\rho$ with $w =1/3$, where $p$ and $\rho$ are the pressure and the energy density, respectively.  This is known as the prethermalization phase~\cite{Berges:2004yj}.

The time it takes for the system to reach chemical and finally thermal equilibrium depends rather sensitively on the input parameters. The Higgs quanta can efficiently  be  produced  via parametric resonance~\cite{Kofman:1994rk} due to the $h^2 \phi^2$ coupling (Fig.~\ref{fig:eos}, right). If the resonance stays active long enough,  chemical equilibrium between the Higgs and inflaton fields sets in earlier. For a substantial $\lambda_h \sim 1$, however, the resonance is shut off by the backreaction effects which induce an extra  contribution to the Higgs mass--squared $\sim \lambda_h \langle h^2 \rangle$. In this case, the Higgs quanta are produced through perturbative scattering and   thermal equilibrium is reached much later.  

The lower bound on the reheating temperature can be estimated by equating   the perturbative interaction rate with the Hubble rate in the radiation--dominated phase. The scattering is expected to be dominated by the $\phi^2 h^2$--interaction, which gives $T_{\rm reh} \gsim
{\cal O}(\lambda_{h \phi}^2)$. For typical coupling values, this results in $T_{\rm reh} \sim 10^{12}$ GeV. 

As the Universe expands and the temperature drops below the inflaton mass, the inflaton 
undegoes the usual ``freeze--out''. Owing to the trilinear $\phi h^2$ interaction, it will quickly decay either into Higgs pairs or light particles (at 1--loop). We emphasize that 
the trilinear term plays a crucial role for consistency of the model: the stable inflaton relics would ``overclose'' the Universe since the  $\phi$--annihilation cross section
is too small to be consistent with the dark matter relic abundance. The latter requires larger couplings, $ \lambda_{h \phi} \sim 10^{-1}-1$ ~\cite{Athron:2017kgt}.


\section{Vacuum stability and low energy constraints}

 
Our next step is to analyze constraints on the model imposed by vacuum stability. 
In this low energy analysis, the dimensionful parameters play a crucial role.

Presently, the curvature is so small that the distinction between the Jordan and Einstein frames becomes immaterial. Thus, we may focus  entirely on the potential $V(\phi,h)$ of Eq.~(\ref{pot}),  treating $\phi$ and $h$ as canonically normalized scalars.
In general, both the Higgs and the inflaton develop  vacuum expectation values 
(VEVs) at the minimum of the potential,     $v\equiv\langle h \rangle$ and $u\equiv\langle\phi \rangle$.
It is convenient, however, to redefine the inflaton field   $\phi^\prime= \phi - u$ such that 
$\langle \phi^\prime \rangle =0$. In terms of the primed field, the potential  
retains the same form (\ref{pot})  if we define the primed \emph{dimensionful} parameters as~\cite{Espinosa:2011ax,Chen:2014ask}
  \beq
\label{eq:shift}
\begin{array}{rcl}
b_3'  &=& b_3 + 3\lif u  \,,\\
\muf^{\prime\,2} &=& \muf^2 - 3\lif u^2 - 2b_3u \,,\\
b_1' &=& b_1 + \lif u^3 + b_3 u^2 - \muf^2u \, \\
\sigma' &=& \sigma + \lhf u\,, \\
\mu_h^{\prime\,2} &=& \mu_h^2 - \lhf u^2 - 2\sigma u\,.
\end{array}
\eeq
 Note that the dimensionless couplings are not affected by this redefinition.
 At the electroweak minimum
$ (\langle h \rangle,\langle\phi^\prime \rangle)=   (v,0)$, the Higgs and the inflaton mix such that the mass eigenstates $h_1, h_2$ are given by 
 \beq
\left(\,\begin{matrix} h_1 \\ h_2\end{matrix}\,\right) = 
\left(\,
\begin{matrix} 
\cos\theta & \sin\theta \\
-\sin\theta & \cos\theta
\end{matrix}
\,\right)
\left(\,\begin{matrix} h-v \\ \phi^\prime\end{matrix}\,\right)\,.
\eeq
The masses $m_{1,2}$  of $h_{1,2}$ and the mixing angle $\theta$ are related to the input parameters by
\beq
\label{eq:constr}
\begin{array}{rcl}
2\lih v^2 &=& m_1^2\cos^2\theta + m_2^2\sin^2\theta\,, \\
\lhf v^2 - \muf^{\prime 2} &=& m_1^2\sin^2\theta + m_2^2\cos^2\theta \,, \\
\sigma^\prime v &=& \dfrac{\sin2\theta}{4}\round{m_1^2 - m_2^2}
\,.  
\end{array}
\eeq
If we identify the observed 125 GeV Higgs--like boson with $h_1$,
  for $m_2>m_1$  the  first relation in  (\ref{eq:constr}) implies that the Higgs self--coupling $\lambda_h$ is greater than that in the SM (obtained by setting $\theta=0$). 
This correction can stabilize the Higgs potential at large field values such that $\lambda_h$
would  never turn negative.  
 
 It is important to note that  a substantial mixing angle $\theta$ implies that 
 $m_2$ cannot be arbitrarily large. Indeed, if $m_2$ is far above the weak scale, the first relation in  (\ref{eq:constr})  makes $\lambda_h$ non--perturbative. In fact, if we require our model to be valid from the electroweak to the Planck (or unitarity) scale, all the mass parameters are confined to the electroweak/TeV scale.

Our next step is to identify parameter regions in which our model remains perturbative up to the Planck scale and the electroweak vacuum remains global. To do that, we use the renormalization group (RG) equations \footnote{We have computed these  equations
analytically and  verified  the result with \texttt{SARAH}~\cite{Staub:2013tta}.}
\bal
\label{eq:rges}
\nn
16\pi^2 {\dfrac{d\lambda_h}{dt}}
&=
24 \lambda_h^2 -6 y_t^4 + \dfrac38
\left( 2 g^4 + (g^2 + g^{\prime 2})^2 \right) 
+ (12 y_t^2 -9 g^2 -3 g^{\prime 2}) \lambda_h + 2 \lhf^2 \;,
\\ \nn
16\pi^2 \dfrac{d\lhf}{dt} 
&=
8 \lhf^2 + 12 \lambda_h \lhf
-\dfrac32 (3 g^2 + g^{\prime 2}) \lhf 
+ 6 y_t^2 \lhf + 6 \lif\lhf \;,
\\ \nn
 16\pi^2 \dfrac{d \lif}{dt} 
 &= 
 8 \lhf^2 + 18 \lif^2 \;,
\\ 
 16\pi^2 \dfrac{d \sigma}{dt} 
 &= 
 \sigma\round{12\lih + 8\lhf - \frac{3g^{\prime 2}}{2} - \frac{9g^2}{2} 
 + 6 y_t^2} + 2\lhf b_3
     \;,
\\ \nn
 16\pi^2 \dfrac{db_3}{dt} 
 &= 
 24\sigma\lhf + 18\lif b_3 \;,
\\ \nn
16\pi^2\dfrac{d y_t}{dt}&= y_t \left(\dfrac92 y_t^2 - \dfrac{17}{12} g'^2 - \dfrac94 g^2 - 8 g_3^2 \right) \;,
\\ \nn
16\pi^2\dfrac{d g_i}{dt}&= c_i \, g_i^3 \quad \textrm{with} \quad (c_1,c_2,c_3)=(41/6,-19/6,-7) \;,
\eal
where  $t= \ln \mu$ with $\mu$ being the energy scale and    $g_i=(g',g,g_3)$ denote the gauge couplings. As the input values at the top quark mass scale $M_t$, we
use $g(M_t)=0.64,\; g'(M_t)=0.35,\; g_3(M_t)=1.16$ and $y_t(M_t)=0.93$. Here we  omit the RG equations for $\mu_i^2$ and $b_1$, which are unimportant for a potential analysis at large field values (although taken into account numerically).

\begin{figure}
\includegraphics[scale=0.5]{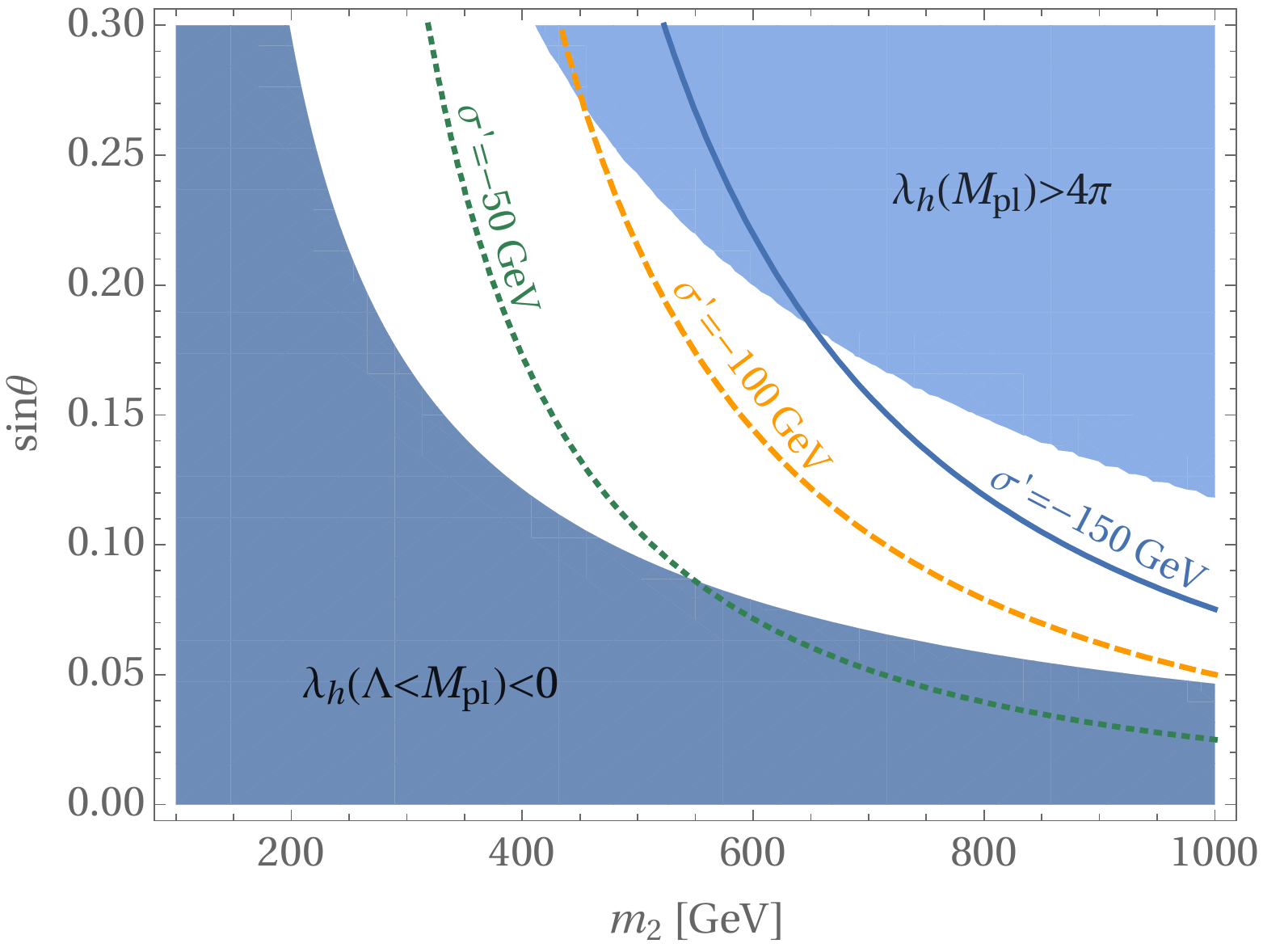}
\hspace{0.05cm}
\includegraphics[scale=0.5]{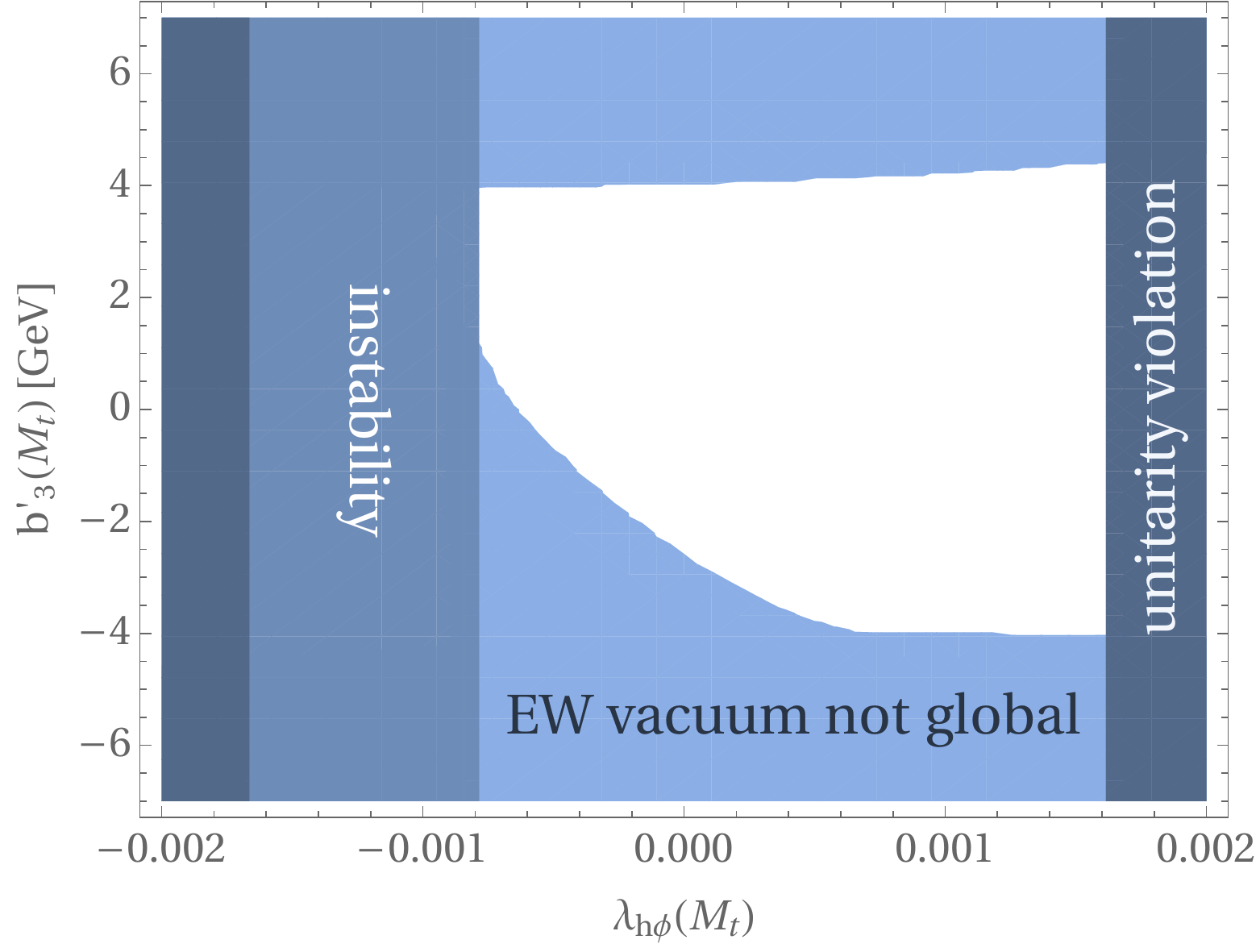}
\caption{\label{fig:space} {\it Left:} Values of $\sin\theta$ and $m_2$ consistent with Higgs potential stability and perturbativity up to $M_{\rm Pl}$ (white region). Also displayed are the curves of constant $\sigma^\prime$. Here $\lambda_{h\phi} = 10^{-3}, \lambda_\phi = 10^{-5}$ at the EW scale and negative $\sin\theta$ are obtained by flipping the sign of $\sigma^\prime$. 
{\it Right:} The $\{\lambda_{h\phi},b_3^\prime \} $  parameter region (in GeV) in which the electroweak vacuum is a global minimum. The other EW scale parameters are fixed to be  $m_2=600 \; {\rm GeV}, \sin\theta=0.144, \lambda_\phi = 10^{-5}$ corresponding to
$\sigma^\prime = -100 $ GeV.  }
\end{figure}

Our results are presented in Fig.~\ref{fig:space}.  
The left panel shows parameter space allowed by perturbativity  
and positivity of $\lambda_h$ at all scales up to $M_{\rm Pl}$. This is analogous to the analysis of 
\cite{Falkowski:2015iwa} for a $Z_2$--symmetric scalar potential. Here we have cut 
$\vert \sin \theta \vert$ at 0.3 which is the upper bound imposed by the Higgs coupling measurements~\cite{TheATLASandCMSCollaborations:2015bln}. (Almost all of the white region with $m_2 > 300$ GeV is also consistent with the LHC and electroweak constraints~\cite{Falkowski:2015iwa,Robens:2016xkb}.) We conclude that 
electroweak to TeV values of $\sigma^\prime $ and $m_2$  
\mk{}
can lead to a stable Higgs potential. 

The right panel shows the  $\{\lambda_{h\phi},b_3^\prime \} $  parameter region in which the electroweak vacuum is the global minimum of the scalar potential. The left part of the panel is excluded by the stability
constraint on the running couplings,
\begin{equation}
\lambda_{h\phi} (\mu) > - \sqrt{\lambda_h(\mu) \lambda_\phi (\mu)} \;, 
\end{equation}
which ensures that there is
no unbounded from below direction at large field values. Relatively large $\vert \lambda_{h \phi}\vert \gsim 2 \times 10^{-3}$ lead to a significant RG contribution to $\lambda_\phi$
 thus violating the unitarity constraint  (\ref{unit})  at the high scale. This excludes the rightmost part of the panel. In the upper and lower shaded regions, there exist further
 minima of the scalar potential at large $\phi^\prime \sim -b_3^\prime/\lambda_\phi$ which are deeper than the electroweak one. 
\mk{We exclude these regions to be conservative although thermal and inflationary effects may stabilize the fields at smaller values in the Early Universe.}

We find that for $u$ up to 10 TeV, the numerical difference between $\sigma$ and $\sigma^\prime$ is negligible. 
In particular, $\sigma \gg \lambda_{h\phi} u$ and 
  according to Eq.~(\ref{eq:constr}) the Higgs--inflaton mixing is governed entirely by the trilinear $\sigma$--term.

We also note that   for negative values of $\lambda_{h\phi}$,    the 
field that drives inflation is a combination of $\phi$ with a  small admixture of $h$~\cite{Lebedev:2011aq}. The Early Universe dynamics develops along the lines discussed above except the reheating process  is expected to be more efficient due to the Higgs interactions.

Our analysis shows that there are exciting prospects for the LHC  new physics searches. First of all, 
the Higgs--inflaton mixing manifests itself as a universal reduction in the Higgs couplings to gauge bosons and fermions. Deviations at  a few percent  level can be detected in the high luminosity LHC phase~\cite{Englert:2014uua}.  Furthermore, the mostly--inflaton state $h_2$ can be found directly as a heavy Higgs--like resonance.  This is facilitated by the decay $h_2 \rightarrow h_1 h_1$ which makes $m_2$ in the TeV range  with $\vert \sin\theta \vert \sim 10^{-1}$ accessible to LHC searches~\cite{Chen:2014ask,Falkowski:2015iwa,Dawson:2017vgm,Lewis:2017dme}.

 
\section{Conclusions}
 
  
We have studied the minimal option of stabilizing the EW vacuum via the Higgs--inflaton mixing, where inflation is driven by a non--minimal scalar coupling to  curvature.
In the presence of the trilinear Higgs--inflaton interaction, such a mixing is inevitable and can significantly  increase the Higgs self--coupling. 
We find that this scenario is cosmologically viable and fits the PLANCK data very well.
The model is particularly attractive when it is described by a single (TeV) mass scale, in which case the mixing angle is substantial. This opens up an exciting avenue for a direct inflaton search at the LHC.

\begin{acknowledgments}
The work of Y.E. was supported in part by JSPS Research Fellowships for Young Scientists and by the Program for Leading Graduate Schools, MEXT, Japan. M.K. is supported by the Academy of Finland project 278722. O.L. and M.Z. acknowledge support from the Academy of Finland project “The Higgs and the Cosmos”.
\end{acknowledgments}

\bibliography{draft}

\end{document}